\documentclass[conference]{IEEEtran}
\IEEEoverridecommandlockouts
\usepackage{cite}
\usepackage{amsmath,amssymb,amsfonts}
\usepackage{graphicx}
\usepackage{textcomp}
\usepackage[dvipsnames]{xcolor}

\usepackage{hyperref}
\usepackage{multirow}
\usepackage{caption}
\usepackage{subcaption}
\usepackage{tabularx}
\usepackage{mathtools}
\usepackage{color}
\usepackage{multicol}
\usepackage{flushend}
\usepackage{listings} 
\usepackage{algorithm}
\usepackage[noend]{algpseudocode}
\usepackage{booktabs}
\usepackage{microtype}
\usepackage[inline]{enumitem}
\usepackage{eso-pic}

\makeatletter
\def\ps@IEEEtitlepagestyle{%
	\def\@oddfoot{\mycopyrightnotice}%
	\def\@oddhead{\hbox{}\@IEEEheaderstyle\leftmark\hfil\thepage}\relax
	\def\@evenhead{\@IEEEheaderstyle\thepage\hfil\leftmark\hbox{}}\relax
	\def\@evenfoot{}%
}
\def\mycopyrightnotice{%
	\begin{minipage}{\textwidth}
		\scriptsize
		\copyright~2024 IEEE. DOI: 10.1109/ICECS61496.2024.10849220.
		Personal use of this material is permitted. 
		Permission from IEEE must be obtained for all other uses,
		in any current or future media, including reprinting/republishing this material
		for advertising or promotional purposes, creating new collective works,
		for resale or redistribution to servers or lists, or reuse of any copyrighted
		component of this work in other works by sending a request to pubs-permissions@ieee.org.
	\end{minipage}
}
\makeatother

\def\BibTeX{{\rm B\kern-.05em{\sc i\kern-.025em b}\kern-.08em
    T\kern-.1667em\lower.7ex\hbox{E}\kern-.125emX}}
\begin{document}
	
\title{\emph{Blink}: Fast Automated Design of Run-Time Power Monitors on FPGA-Based Computing Platforms}

\author{
	\IEEEauthorblockN{Andrea Galimberti}
	\IEEEauthorblockA{\textit{DEIB}\\
			\textit{Politecnico di Milano}\\
			Milano, Italy\\
			andrea.galimberti@polimi.it}
	\and
	\IEEEauthorblockN{Michele Piccoli}
	\IEEEauthorblockA{\textit{DEIB}\\
			\textit{Politecnico di Milano}\\
			Milano, Italy\\
			michele.piccoli@polimi.it}
	\and
	\IEEEauthorblockN{Davide Zoni}
	\IEEEauthorblockA{\textit{DEIB}\\
			\textit{Politecnico di Milano}\\
			Milano, Italy\\
			davide.zoni@polimi.it}
}

\maketitle

\IEEEpeerreviewmaketitle

\begin{abstract}
The current over-provisioned heterogeneous multi-cores require
effective run-time optimization strategies,
and the run-time power monitoring subsystem is paramount for their success.
Several state-of-the-art methodologies address the design of
a run-time power monitoring infrastructure for generic computing platforms.
However, the power model's training requires time-consuming gate-level simulations that,
coupled with the ever-increasing complexity of the modern heterogeneous platforms,
dramatically hinder the usability of such solutions.
This paper introduces \emph{Blink}, a scalable framework for
the fast and automated design of run-time power monitoring infrastructures
targeting computing platforms implemented on FPGA.
\emph{Blink} optimizes the time-to-solution to deliver
the run-time power monitoring infrastructure by
replacing traditional methodologies' gate-level simulations and power trace computations with
behavioral simulations and direct power trace measurements.
Applying \emph{Blink} to multiple designs mixing a set of
HLS-generated accelerators from a state-of-the-art benchmark suite
demonstrates an average time-to-solution speedup of 18 times
without affecting the quality of the run-time power estimates.
\end{abstract}


\section{Introduction}
\label{sec:introduction}
The evolution of computing platforms towards
over-provisioned heterogeneous and multi-core architectures
has made run-time optimizations crucial to maximize energy efficiency.
Knowledge of the dynamic power consumption of the computing platform
during its operation is at the core of any run-time optimization methodology,
and the collection of such information is generally achieved
through an ad-hoc monitoring infrastructure.

The open literature includes a variety of run-time power monitoring frameworks
that augment the existing RTL description of the computing platform by
instrumenting an on-chip hardware component devoted to
the estimation of the dynamic power consumption of the system at run time~\cite{Zoni_2023CSUR}.
State-of-the-art methodologies leverage gate-level simulations of
the target computing platform to extract a set of estimated power traces and
the corresponding switching activity that are used to identify a power model.
The identified power model is then processed to generate an RTL description that 
meets the accuracy, temporal resolution, and area requirements and is 
finally instrumented into the computing platform,
thus delivering the run-time power monitoring infrastructure.

On the model family side, linear regression models
are most commonly used due to their low complexity and overheads
while producing estimation errors below 5\%~\cite{Najem_TCAD2017}.
Machine-learning-based solutions, on the contrary, result instead in
more significant overheads due to the additional complexity without
outpacing linear regression models in accuracy~\cite{Zhe_FPL2017}.

The literature includes several solutions to monitor the platform's power consumption
at run time.
Software-implemented power monitors leverage the relationship
between power consumption and performance monitoring counters and
require no microarchitectural modifications,
affecting however the monitored system's performance
due to monitoring being performed as
a software application~\cite{Rodrigues_TCAS-II2013,Walker_TCAD2017}.
In contrast, hardware-based ones offer high accuracy
with no performance overhead by employing
an ad-hoc hardware infrastructure to monitor
selected signals' switching activity and deliver the power estimates,
which results in area overhead~\cite{Zhe_FPL2017,Xie_ICCAD2022}. 

All the run-time power monitoring frameworks from the literature
notably rely on time-consuming gate-level simulations to
identify their underlying model, thus they are only suited to
small computing platforms due to their lengthy execution times~\cite{Zoni_2023CSUR}.
Conversely, the ever-increasing complexity of the computing platforms further widens
the gap between the tight time-to-market deadlines and the lengthy design process,
thus motivating the investigation of novel techniques.

\begin{figure*}
	\begin{minipage}[b]{\textwidth}
		\centering
		\begin{subfigure}[t]{0.92\textwidth}
			\centering
			\includegraphics[width=\textwidth]{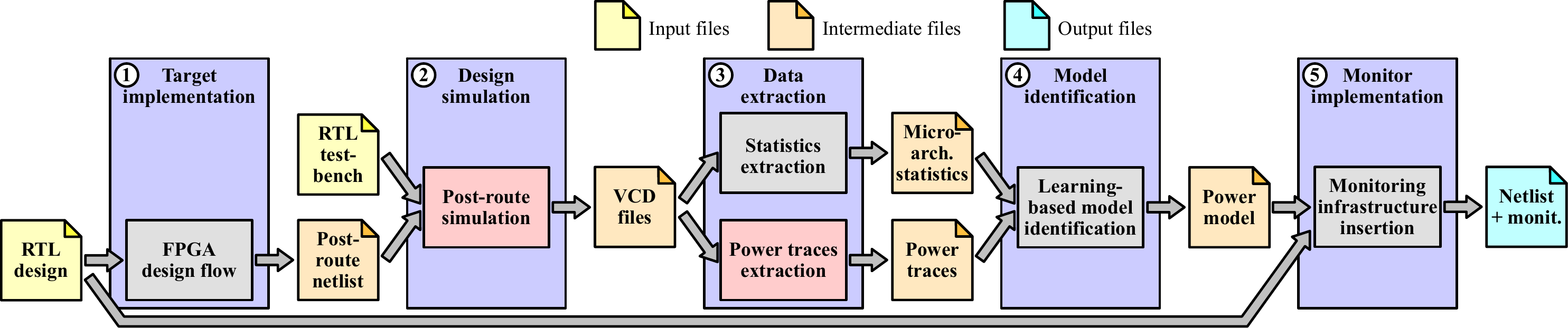}
			\caption{State-of-the-art simulation-based framework~\cite{Zoni_IGSC2020_PowerModel}}
			\label{fig:flowchart_baseline}
		\end{subfigure}

		\vspace{0.16cm} % Add vertical space here
		
		\begin{subfigure}[b]{0.92\textwidth}
			\centering
			\includegraphics[width=\textwidth]{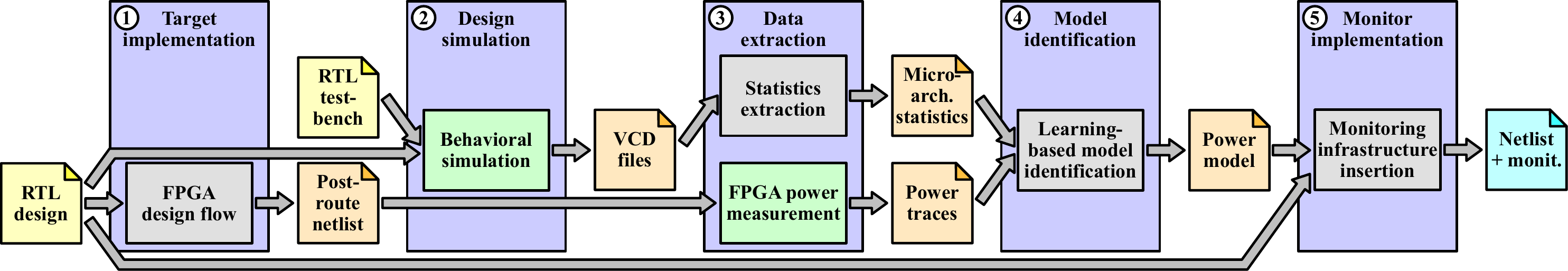}
			\caption{\emph{Blink} framework}
			\label{fig:flowchart_proposed}
		\end{subfigure}
	\end{minipage}
	\caption{Flowchart diagrams of the (a) state-of-the-art and (b) \emph{Blink} frameworks,
		with the most time-consuming steps of the state-of-the-art framework in \textcolor{BrickRed}{\textbf{red}}
		and \emph{Blink}'s improvements in \textcolor{ForestGreen}{\textbf{green}}.}
	\label{fig:flowchart}
\end{figure*}

\paragraph*{Contributions}
This manuscript introduces \emph{Blink}, a novel methodology that leverages
a hybrid simulation-measurement approach to speed up
the automated instrumentation of an ad-hoc run-time power monitoring infrastructure
into generic computing platforms targeting deployment on FPGAs.

\emph{Blink} dramatically reduces the time-to-solution,
compared to state-of-the-art simulation-based methodologies,
by replacing their lengthy gate-level simulations to collect
the switching activity and estimated power traces with
\begin{enumerate*}[label=\textit{\roman*)}]
	\item fast behavioral simulations to collect the switching activity and
	\item direct measurements of the power traces straight from the FPGA.
\end{enumerate*}

Thanks to its speed, \emph{Blink} provides a scalable framework to design
and instrument run-time power monitoring infrastructures,
outperforming state-of-the-art solutions by 18\texttimes\, on average
with comparable area and power overheads and estimation accuracy
in our experimental evaluation.

\section{Methodology}
\label{sec:methodology}
This section describes the \emph{Blink} methodology by highlighting
the key changes to the state-of-the-art run-time power monitoring methodologies
that allow it to deliver a dramatic time-to-solution speedup with comparable accuracy and overheads.

\subsection{State-of-the-art framework}
A state-of-the-art simulation-based framework~\cite{Zoni_IGSC2020_PowerModel}
produces the gate-level netlist of the target computing platform
enhanced with run-time power monitoring capabilities
from its RTL description and the corresponding testbench. 
As shown in \figurename~\ref{fig:flowchart_baseline}, it can be split into five phases:

\subsubsection{Target implementation} The standard hardware design flow implements
the target computing platform by taking its RTL description and performing
synthesis, place, and route to produce the corresponding gate-level netlist.

\subsubsection{Design simulation} A gate-level simulation stresses
each part of the post-route netlist of the target computing platform
to produce a value change dump~(VCD) file that contains
the statistics needed to compute the power consumption trace and
the corresponding switching activity.
The time required by the simulation notably grows as
the size of the design increases.

\subsubsection{Data extraction} This phase processes the VCD file to
extract the switching activity and compute the estimated power traces.
The latter is notably a lengthy task whose execution time increases as
the complexity of the computing platform grows and
the temporal resolution of the power traces shrinks.

\subsubsection{Model identification} A power model, consisting of
a subset of the signals in the target computing platform and
their corresponding weights, is identified by correlating
the switching activity and the power trace,
minimizing the distance between
the actual power trace and the one produced by
the identified power model subject to additional requirements,
e.g., the maximum number of signals selected as model inputs.

\subsubsection{Monitor implementation} A hardware monitoring infrastructure
is automatically instrumented into the original target computing platform,
wrapping signals identified by the model to periodically
read out their switching activity and compute the power estimate and exposing the latter
through a register. 

\begin{table*}[t]
	\centering
	\caption{Accelerator designs and corresponding monitoring infrastructure.
		Legend: \textbf{Freq} clock frequency~(MHz), \textbf{Pwr} power consumption~(W),
		\textbf{HW} Hamming-weight and \textbf{ST} single-toggle counters,
		\textbf{Acc} accuracy,
		\checkmark yes, \texttimes no.}
	\begin{tabular}{cccccrrrrrrrrrrrr}
		\toprule
		                                                                                                  \multicolumn{5}{c}{\textbf{Accelerator design configuration}}                                                                                                   &  \multicolumn{4}{c}{\textbf{FPGA resource utilization}}   &               &              & \multicolumn{2}{c}{\textbf{Counters}} &   \multicolumn{3}{c}{\textbf{Overhead}}   & \multicolumn{1}{c}{\textbf{Acc}} \\
		\cmidrule(lr){1-5} \cmidrule(lr){6-9} \cmidrule(lr){12-13} \cmidrule(lr){14-16} \cmidrule(lr){17-17}
		                                                                             \textbf{ID} & \textbf{AES} & \textbf{Blowfish} & \textbf{GSM} & \textbf{MIPS} & \textbf{LUT} & \textbf{FF} & \textbf{BRAM} & \textbf{DSP} & \textbf{Freq} & \textbf{Pwr} & \textbf{HW} &             \textbf{ST} & \textbf{LUT} & \textbf{FF} & \textbf{Pwr} &                    \textbf{RMSE} \\ \midrule
		                                                                                           \emph{A1}                                                                                            &  \checkmark  &    \texttimes     &  \texttimes  &  \texttimes   &        13764 &       14674 &             0 &            0 &           150 &         0.22 &           4 &                       0 &        2.8\% &       1.9\% &        1.6\% &                            4.3\% \\
		                                                                                           \emph{A2}                                                                                            &  \checkmark  &    \texttimes     &  \texttimes  &  \texttimes   &        24412 &       26170 &             0 &            0 &           150 &         0.47 &           5 &                       0 &        2.1\% &       1.7\% &        1.2\% &                            4.5\% \\
		                                                                                           \emph{A3}                                                                                            &  \checkmark  &    \checkmark     &  \texttimes  &  \texttimes   &        50408 &       72307 &          66.5 &            0 &           100 &         0.50 &           6 &                       0 &        1.0\% &       0.9\% &        1.0\% &                            4.1\% \\
		                                                                                           \emph{A4}                                                                                            &  \checkmark  &    \texttimes     &  \texttimes  &  \checkmark   &        41067 &       29802 &            30 &          160 &           130 &         0.37 &           7 &                       0 &        2.1\% &       2.1\% &        1.8\% &                            3.6\% \\
		                                                                                           \emph{A5}                                                                                            &  \texttimes  &    \checkmark     &  \texttimes  &  \texttimes   &        45793 &       72098 &          92.5 &            0 &           100 &         0.32 &           4 &                       2 &        0.5\% &       0.4\% &        0.8\% &                            4.3\% \\
		                                                                                           \emph{A6}                                                                                            &  \texttimes  &    \texttimes     &  \texttimes  &  \checkmark   &        43325 &       23814 &            30 &          240 &           120 &         0.24 &           8 &                       0 &        1.2\% &       2.6\% &        2.9\% &                            4.9\% \\
		                                                                                           \emph{A7}                                                                                            &  \texttimes  &    \checkmark     &  \texttimes  &  \checkmark   &        32931 &       36907 &          49.5 &           80 &           100 &         0.18 &           8 &                       0 &        0.8\% &       1.0\% &        0.4\% &                            4.9\% \\
		                                                                                           \emph{A8}                                                                                            &  \texttimes  &    \texttimes     &  \checkmark  &  \checkmark   &        25216 &       23572 &             0 &          240 &           100 &         0.19 &           3 &                       1 &        0.7\% &       0.9\% &        0.7\% &                            3.4\% \\
		                                                                                           \emph{A9}                                                                                            &  \texttimes  &    \checkmark     &  \checkmark  &  \checkmark   &        48226 &       60573 &            75 &          240 &           100 &         0.29 &           5 &                       0 &        0.7\% &       0.7\% &        0.6\% &                            4.4\% \\
		                                                                                          \emph{A10}                                                                                            &  \checkmark  &    \checkmark     &  \checkmark  &  \checkmark   &        53500 &       66373 &            75 &          240 &           100 &         0.40 &           9 &                       1 &        1.9\% &       1.4\% &        0.1\% &                            3.9\% \\ \bottomrule
	\end{tabular}
	\label{tab:soc_overhead_accuracy}
\end{table*}

\subsection{Blink framework}
The \emph{Blink} methodology, depicted in \figurename~\ref{fig:flowchart_proposed},
improves the state-of-the-art frameworks by optimizing their most time-consuming phases,
i.e., \emph{design simulation} and \emph{data extraction}.
In particular, it replaces the post-route simulation and power traces extraction steps,
in red in \figurename~\ref{fig:flowchart_baseline},
with the corresponding and significantly faster behavioral simulation and FPGA power measurement steps,
in green in \figurename~\ref{fig:flowchart_proposed}.

\subsubsection*{FPGA power measurement}
\emph{Blink} substitutes the lengthy extraction of the estimated power traces
in the \emph{data extraction} phase, that takes the longest in
state-of-the-art frameworks, with the direct measurement of
power consumption from an FPGA prototype board that hosts the target computing platform.
A bitstream is first obtained from the post-route netlist of the latter and flashed on the FPGA,
then a power trace is collected directly from the FPGA protoype board
through an oscilloscope while the computing platform is running.
The accurate temporal alignment between the VCD obtained from the \emph{design simulation} and
the measured power trace measured is notably paramount to enabling an effective identification of
the run-time power model of the target computing platform.
A hardware trigger pin enables such alignment
and the adoption of a delay before the actual measurement minimizes
power distortion due to ringing phenomena.

\subsubsection*{Behavioral simulation}
The direct measurement of the power traces, in place of their extraction from VCD files,
makes it possible to also optimize the \emph{design simulation} phase.
The \emph{Blink} methodology does not require indeed a gate-level simulation of
the post-route netlist of the target computing platform, but it can instead
run a simpler and faster behavioral simulation of the RTL design to
obtain a VCD that can be used solely for the extraction of
the microarchitectural statistics, i.e., the switching activity.

The other three phases are instead the same as the state-of-the-art simulation-based framework ones,
with only few minor changes to support \emph{Blink}'s improvements.
In particular, the \emph{model identification} step takes
the measured power consumption trace as an input rather than
the estimated power trace extracted from the gate-level switching activity,
which is equivalent from the execution time standpoint.

\begin{table*}[t]
	\centering
	\caption{Execution time (in minutes) of the reference state-of-the-art and
		\emph{Blink} run-time power monitoring frameworks, applied to
		the accelerator designs described in \tablename~\ref{tab:soc_overhead_accuracy}.
		Legend: \textbf{DUT} design under test, \textbf{SOTA} state-of-the-art.}
	\begin{tabular}{crrrrrrrrrrrrr}
		\toprule
		                                                           \multicolumn{1}{c}{\textbf{DUT}}                                                            & \multicolumn{2}{c}{\textbf{Target impl.}} & \multicolumn{2}{c}{\textbf{Design sim.}} & \multicolumn{2}{c}{\textbf{Data extr.}} & \multicolumn{2}{c}{\textbf{Model ident.}} & \multicolumn{2}{c}{\textbf{Monitor impl.}} & \,\,\,\,\,\,\,\,\,\, &  \multicolumn{2}{c}{\textbf{Total}}   \\
		\cmidrule(lr){1-1} \cmidrule(lr){2-3} \cmidrule(lr){4-5} \cmidrule(lr){6-7} \cmidrule(lr){8-9} \cmidrule(lr){10-11} \cmidrule(lr){13-14}
		\textbf{ID} & \textbf{SOTA} &     \textbf{\emph{Blink}} & \textbf{SOTA} &    \textbf{\emph{Blink}} & \textbf{SOTA} &   \textbf{\emph{Blink}} & \textbf{SOTA} &     \textbf{\emph{Blink}} & \textbf{SOTA} &      \textbf{\emph{Blink}} &                      & \textbf{SOTA} & \textbf{\emph{Blink}} \\ \midrule
		                                                                      \emph{A1}                                                                        &            11 &                        11 &           781 &                       34 &           912 &                      11 &             4 &                         5 &            12 &                         13 &                      &          1720 &                    74 \\
		                                                                      \emph{A2}                                                                        &            16 &                        16 &           804 &                       50 &          1189 &                      16 &             5 &                         4 &            18 &                         17 &                      &          2032 &                   103 \\
		                                                                      \emph{A3}                                                                        &            28 &                        28 &          1474 &                      116 &          2461 &                      14 &             3 &                         4 &            29 &                         30 &                      &          3995 &                   192 \\
		                                                                      \emph{A4}                                                                        &            33 &                        33 &          1411 &                       92 &          1879 &                      16 &             5 &                         5 &            34 &                         34 &                      &          3362 &                   180 \\
		                                                                      \emph{A5}                                                                        &            26 &                        26 &          1474 &                       80 &          1911 &                       8 &             4 &                         4 &            27 &                         28 &                      &          3442 &                   146 \\
		                                                                      \emph{A6}                                                                        &            19 &                        19 &           268 &                       16 &           305 &                       4 &             5 &                         5 &            21 &                         21 &                      &           618 &                    65 \\
		                                                                      \emph{A7}                                                                        &            21 &                        21 &          1131 &                       86 &          1196 &                       6 &             5 &                         4 &            22 &                         23 &                      &          2375 &                   140 \\
		                                                                      \emph{A8}                                                                        &            15 &                        15 &           370 &                       37 &           501 &                       6 &             4 &                         5 &            17 &                         18 &                      &           907 &                    81 \\
		                                                                      \emph{A9}                                                                        &            27 &                        27 &          1519 &                      120 &          2025 &                       8 &             5 &                         4 &            29 &                         28 &                      &          3605 &                   187 \\
		                                                                      \emph{A10}                                                                       &            32 &                        32 &          1702 &                      131 &          2061 &                      10 &             5 &                         5 &            34 &                         33 &                      &          3834 &                   211 \\ \midrule
		                                                                   \textbf{Speedup}                                                                    &        \multicolumn{2}{r}{1.00\texttimes} &      \multicolumn{2}{r}{15.11\texttimes} &    \multicolumn{2}{r}{150.76\texttimes} &        \multicolumn{2}{r}{1.01\texttimes} &         \multicolumn{2}{r}{0.99\texttimes} &                      &   \multicolumn{2}{r}{18.12\texttimes} \\ \bottomrule
	\end{tabular}
	\label{tab:soc_speedup}
\end{table*}

\section{Experimental evaluation}
\label{sec:experiments}
\begin{figure}
	\centering
	\includegraphics[width=0.8\columnwidth]{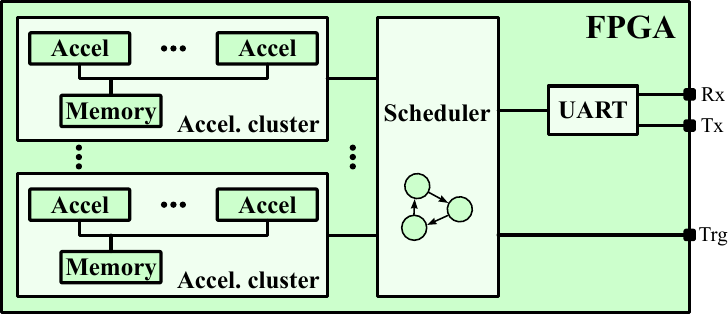}
	\caption{Architecture of the benchmarking platform
		for the accelerator designs targeted by the experimental campaign.}
	\label{fig:bench_arch}
\end{figure}

The experimental campaign evaluates the area and power overheads and accuracy
of run-time monitoring infrastructures obtained by applying \emph{Blink}
as well as the time-to-solution speedup compared to state-of-the-art simulation-based frameworks.
The experiments target a benchmarking platform,
depicted in \figurename~\ref{fig:bench_arch},
that instantiates a set of accelerator designs.
The host PC drives the computation by interacting
via UART (Rx and Tx in \figurename~\ref{fig:bench_arch}),
while a trigger signal~(Trg) marks the start and end of
the computation for the oscilloscopes and
is emulated when simulating the target computing platform
so that the VCDs accurately match the measured power traces.

The accelerator architecture instantiates accelerator clusters that are
enabled by a scheduler to have a variable number of active resources at a time.
Each cluster contains a number of accelerators of the same type that
share a memory for their input and output data.
Multiple clusters in the same design can include different types of accelerators.
The latter are obtained through high-level synthesis~(HLS) of
the \emph{AES}, \emph{Blowfish}, \emph{GSM}, and \emph{MIPS} applications from
the \emph{CHStone} benchmark suite~\cite{Hara_ISCAS2008}.

The hardware flow targets an \emph{AMD Artix-7 100} FPGA
mounted on a \emph{NewAE Technology CW305 Artix FPGA Target} board
and the \emph{AMD Vitis 2023.1} toolchain is employed for HLS,
RTL synthesis, place-and-route, bitstream generation, and simulation.
Power consumption is computed
from the voltage drop measured by
two \emph{Pico Technology PicoScope 5244D} oscilloscopes
across a 100m$\Omega$ shunt resistor.

The identification of the power model and
the evaluation of power estimates obtained by the monitoring infrastructure
are carried out by splitting the collected dataset according to a 80:20 ratio and
targeting a 10\textmu s temporal resolution.
Model identification and monitoring implementation are performed
according to the state-of-the-art simulation-based methodology
selected as a reference~\cite{Zoni_IGSC2020_PowerModel}
to provide the fairest comparison.

The first-order linear model takes as its inputs the switching activity of
a selected subset of input and output signals of the modules in the design hierarchy
and outputs the estimated power consumption.
The hardware counters that monitor the selected signals are either single-toggle ones,
counting any change in the target multi-bit signal, or Hamming-weight counters,
that count the number of bits that toggled their values.

\subsection{Overhead and accuracy results}
\label{ssec:exp_overhead_accuracy}
We evaluate the area overhead of the monitoring infrastructures
instantiated by \emph{Blink}
as the number of additional FPGA resources required to
implement the hardware counters and compute the power estimate,
and their accuracy as the root-mean-square error~(RMSE) of the estimation.
\emph{Blink}'s measures of the power consumption of the actual target computing platform
allow us to provide accurate data also with respect to the power overhead of the monitoring infrastructure,
that is often neglected or inaccurate in the literature~\cite{Zoni_2023CSUR}.

The experiments target designs with
various combinations of the accelerators and
clusters containing different numbers of the latter.
\tablename~\ref{tab:soc_overhead_accuracy} details
the instantiated accelerators,
the FPGA resource utilization, the clock frequency targeted by synthesis and place-and-route, and
the maximum power consumption of ten accelerator designs.
The latter
target clock frequencies in a 100-150MHz range and
show a maximum power consumption between 0.18W and 0.50W.
\tablename~\ref{tab:soc_overhead_accuracy} also lists the quality metrics
related to the monitoring infrastructure, reporting
the number of single-toggle and Hamming-weight counters that monitor
the toggling activity of selected signals as well as the LUT and FF area overhead,
the power overhead, and the RMSE estimation accuracy metric.
The results highlight area and power overheads below 3\% and
an RMSE below 5\%, on par with the state-of-the-art methodologies~\cite{Zoni_2023CSUR}.

\subsection{Time-to-solution speedup}
\label{ssec:exp_speedup}
\tablename~\ref{tab:soc_speedup} provides a breakdown of the execution times of
the five framework phases for the application of
the state-of-the-art~\cite{Zoni_IGSC2020_PowerModel} and \emph{Blink} methodologies
to the ten accelerator design instances, and its bottom row lists
the average speedup for each phase as well as for their total.

\emph{Blink}'s execution times are notably in the order of few hours,
ranging from around 1 to less than 4 hours, while the state-of-the-art methodology
requires tens of hours, with execution times ranging instead from 10 to more than 66 hours.
Whereas it took almost 18 days overall to obtain a monitoring-enhanced netlist
for each of the accelerator designs by using the state-of-the-art flow,
applying \emph{Blink} required slightly less than 1 day,
with an overall time-to-solution speedup in a range comprised between
9\texttimes\, and 23\texttimes, and more than 18\texttimes\, on average.
Design simulation and data extraction are the phases with the largest time savings,
with speedups of 15\texttimes\, and 151\texttimes, respectively.

Employing a different commercial simulator might reduce the execution time of
the post-route simulation in the state-of-the-art framework,
but would remarkably not avoid the need to perform the computation of the power trace,
which mandates instead the usage of the \emph{AMD Vitis} toolchain to obtain
a power trace that is accurate enough for modeling purposes.

\section{Conclusions}
\label{sec:conclusions}
This paper introduced \emph{Blink}, a scalable framework for
the fast and automated design of run-time power monitoring infrastructures
targeting FPGA-based computing platforms
that optimizes time-to-solution by replacing
the lengthy gate-level simulations and power trace computations used in traditional methodologies with
faster behavioral simulations and direct power trace measurements.
Applying \emph{Blink} to various realistic accelerator-based designs
demonstrated an average time-to-solution speedup of 18\texttimes\,
with comparable overheads and accuracy.
Future extensions foresee its application to RISC-V-based systems-on-chip~\cite{Zoni_2022JSA}
and cryptography accelerators~\cite{Galimberti_2022DSD}.

\bibliographystyle{IEEEtran}
\bibliography{IEEEabrv,2024_icecs}

\end{document}